\documentclass{article}

\usepackage{arxiv}

\usepackage[utf8]{inputenc} 
\usepackage[T1]{fontenc}    
\usepackage{hyperref}       
\usepackage{url}            
\usepackage{booktabs}       
\usepackage{amsfonts}       
\usepackage{nicefrac}       
\usepackage{microtype}      
\usepackage{lipsum}
\usepackage{graphicx}
\usepackage{amsmath}
\usepackage{breqn}
\graphicspath{ {./images/} }

\title{Full waveform inversion with CNN-based velocity representation extension}

\author{
 Xinru Mu \\
  Physical Science and Engineering Division\\
  King Abdullah University of Science and Technology\\
  Thuwal 23955, Saudi Arabia \\
  \texttt{xinru.mu@kaust.edu.sa} \\
   \And
Omar M. Saad \\
  Physical Science and Engineering Division\\
  King Abdullah University of Science and Technology\\
  Thuwal 23955, Saudi Arabia \\
  \And
 Tariq Alkhalifah \\
  Physical Science and Engineering Division\\
  King Abdullah University of Science and Technology\\
  Thuwal 23955, Saudi Arabia \\
}

\begin{document}
\maketitle

\begin{abstract}
Full waveform inversion (FWI) updates the velocity model by minimizing the discrepancy between observed and simulated data. However, discretization errors in numerical modeling and incomplete seismic data acquisition can introduce noise, which propagates through the adjoint operator and affects the accuracy of the velocity gradient, thereby impacting the FWI inversion accuracy. To mitigate the influence of noise on the gradient, we employ a convolutional neural network (CNN) to refine the velocity model before performing the forward simulation, aiming to reduce noise and provide a more accurate velocity update direction. We use the same data misfit loss to update both the velocity and network parameters, thereby forming a self-supervised learning procedure. We propose two implementation schemes, which differ in whether the velocity update passes through the CNN. In both methodologies, the velocity representation is extended (VRE) by using a neural network in addition to the grid-based velocities. Thus, we refer to this general approach as VRE-FWI. Synthetic and real data tests demonstrate that the proposed VRE-FWI achieves higher velocity inversion accuracy compared to traditional FWI, at a marginal additional computational cost of approximately 1\%.
\end{abstract}


\section{Introduction}
Seismic velocity inversion is essential for subsurface imaging and hydrocarbon reservoir prediction. Full Waveform Inversion (FWI) is a high-resolution velocity inversion method that updates the velocity model by minimizing the residuals between observed and simulated data \cite{tarantola1984inversion, virieux2009overview}. However, FWI is a highly nonlinear and ill-posed problem, making it sensitive to the initial model, noise, and multiparameter crosstalk. These challenges can lead the inversion process to converge to a local minimum. For instance, FWI is particularly susceptible to numerical simulation errors and incomplete data acquisition, which can introduce noise into the gradient and degrade the accuracy of velocity inversion. To mitigate gradient noise, common approaches include direct smoothing and Tikhonov regularization \cite{aghazade2023full}. However, these methods inevitably modify the gradient, potentially compromising the data-matching objective.

Deep neural networks (DNNs), as powerful function approximators, can effectively learn the nonlinear mapping relationships within high-dimensional complex data \cite{hornik1989multilayer, lecun2015deep}. They have been successfully applied in various seismic processing tasks, including denoising, interpolation, velocity inversion, imaging, and fault interpretation, demonstrating remarkable performance \cite{wang2019deep, wu2019faultseg3d, wu2019parametric, li2021deep, liu2022efficient, saad2024unsupervised, cheng2024meta, wang2024enhanced, wu2024least}. In seismic velocity inversion, geophysicists employ data-driven approaches to learn the nonlinear relationship between observed data and velocity models \cite{yang2019deep, wu2019inversionnet, liu2021deep, kazei2021mapping}. A trained neural network is then used to directly predict the velocity model from the observed data. Some data-driven regularization methods are frequently used to improve the accuracy and stability of FWI inversion \cite{zhang2022regularized, wang2023prior, li2024mau}. In addition, DNNs have been used to construct misfit functions that enhance the measurement of differences between observed and simulated data \cite{sun2022ml}, thereby further alleviating the local minima problem and improving inversion accuracy. However, these data-driven methods require a large amount of training data and substantial computational resources for model training. Moreover, obtaining labeled data for real-world applications is difficult, and models trained on synthetic data often fail to generalize well to real seismic datasets due to the lack of physical constraints, resulting in unreliable results \cite{sun2022ml, yang2023fwigan}. 

Using self-supervised or unsupervised machine learning methods to enhance the accuracy of FWI presents a promising avenue with significant potential for practical applications. \cite{yang2023fwigan} introduced a physics-informed generative adversarial network to measure the loss in FWI in an unsupervised manner, which reduces the risk of getting trapped in local minima. \cite{saad2024siamesefwi} proposed a Siamese network to transform seismic data into a latent space, allowing for better data comparison and consequently enhancing inversion accuracy. Additionally, deep-image-prior FWI (FWIDIP) has been proposed to improve the accuracy of FWI inversion \cite{wu2019parametric, he2021reparameterized, zhu2022integrating}. This method employs a neural network to represent the velocity model and updates its parameters to minimize the discrepancy between observed and simulated data. \cite{dhara2023elastic} utilized a deep convolutional encoder-decoder network to represent P-wave and S-wave velocities, along with density, effectively mitigating parameter crosstalk and providing high-accuracy results for multi-parameter inversion. \cite{sun2023implicit} employed Implicit Neural Representation (INR) to represent velocity models across various grid scales. The FWIDIP and INR-FWI methods, which fully replace the grid-based velocity representation with a neural network, can enhance inversion accuracy. However, due to the inherent frequency bias, they also introduce substantial computational cost, typically requiring thousands of iterations. Moreover, for complex velocity models, larger neural networks are often needed, resulting in increased memory usage.

In this work, we propose a self-supervised learning framework to suppress gradient noise arising from the numerical implementation of FWI and insufficient data sampling coverage, thereby enhancing the accuracy of velocity inversion. We employ a convolutional neural network (CNN) to refine the velocity model, which is then used for forward modeling to generate simulated data. The residual between the simulated and observed data is used for backpropagation to update both the velocity model and the network parameters. Based on this fundamental principle, we propose two implementation frameworks that differ in whether the CNN is incorporated into the backpropagation process for updating the velocity model. Given that the CNN functions as an extension of the grid-based velocity representation, we refer to this method as velocity representation extension FWI (VRE-FWI). Testing on two synthetic datasets and one real dataset demonstrates that the proposed VRE-FWI effectively suppresses gradient noise compared to traditional FWI, resulting in more accurate gradient updates and velocity inversion results.

\section{Theory}
Full waveform inversion relies heavily on the velocity representation. Therefore, instead of replacing the velocity model or its updates with a neural network, we adopt a complementary approach wherein the neural network supports the conventional grid representation. This form of representation allows us to maintain FWI efficiency while benefiting from the network to identify and filter out information that might not support the data fitting objective. In other words, we extend the conventional gridded representation with additional neural network parameters. The key here is that we update both the velocity and the neural network and we hope that the neural network can, in one implementation, absorb all noise and artifacts as it supports the data fitting objective, and in the other approach, learn to be a filter of these noise and artifacts.

\subsection{Review of acoustic FWI}
The 2D constant-density acoustic wave equation is expressed as follows:

\begin{dmath}
\label{eq1}
\frac{{{\partial ^2}u(\textbf{x},t)}}{{\partial {t^2}}} = {v^2}(\textbf{x})\left( {\frac{{{\partial ^2}u(\textbf{x},t)}}{{\partial {x^2}}} + \frac{{{\partial ^2}u(\textbf{x},t)}}{{\partial {z^2}}}} \right) + f({\textbf{x}_s},t)\delta (\textbf{x} - {\textbf{x}_s}),
\end{dmath}
where \({u(\textbf{x},t)}\) is the displacement field, \(v(\textbf{x})\) is the velocity, \({\bf{x}} = \left\{ {x,z} \right\}\) is the spatial coordinate vector, \(t\) denotes time, \(f({\textbf{x}_s},t)\) is the source time function, \({\textbf{x}_s}\) is the source location, \(\delta \) represents the Dirac delta function. We solve the acoustic wave equation (\ref{eq1}) using a finite-difference scheme with second-order accuracy in time and eighth-order accuracy in space.

We aim to recover a subsurface velocity model based on the following Euclidean distance: 
\begin{equation}
\label{eq2}
J({\bf{m}}) = {\left\| {{{\bf{d}}_{{\rm{cal}}}}({\bf{m}}) - {{\bf{d}}_{{\rm{obs}}}}} \right\|_2},
\end{equation}
where \({{{\bf{d}}_{{\rm{cal}}}}}\) and \({{{\bf{d}}_{{\rm{obs}}}}}\) are the calculated and observed data, respectively, \(\textbf{m}\) is a representation of the subsurface velocity model, like velocity, or slowness squared. FWI often relies on the following velocity update formula to iteratively minimize the difference between the observed and simulated data:
\begin{equation}
\label{eq2}
\textbf{m} = \textbf{m} + \lambda  * \frac{{\partial J}}{{\partial \textbf{m}}},
\end{equation}
where \(\lambda\) is the step length and \(\frac{{\partial J}}{{\partial \textbf{m}}}\) is the gradient with respect to the velocity model. We usually use the adjoint state method to compute the gradient with respect to the velocity \cite{plessix2006review}. In this study, we use the Deepwave toolbox \cite{richardson_alan_2023} within the PyTorch environment to implement all FWI processes. The Deepwave toolbox constructs FWI as a recurrent neural network (RNN) and implements it using deep learning software such as PyTorch. Compared to traditional FWI, the velocity gradients are computed using automatic differentiation, which yields results approximately equal to those obtained through the traditional adjoint-state method \cite{richardson2018seismic, sun2020theory}. Additionally, this RNN-based approach facilitates the integration of deep learning algorithms.

\subsection{The proposed VRE-FWI}
\begin{figure*}
\centering
\includegraphics[width=\textwidth]{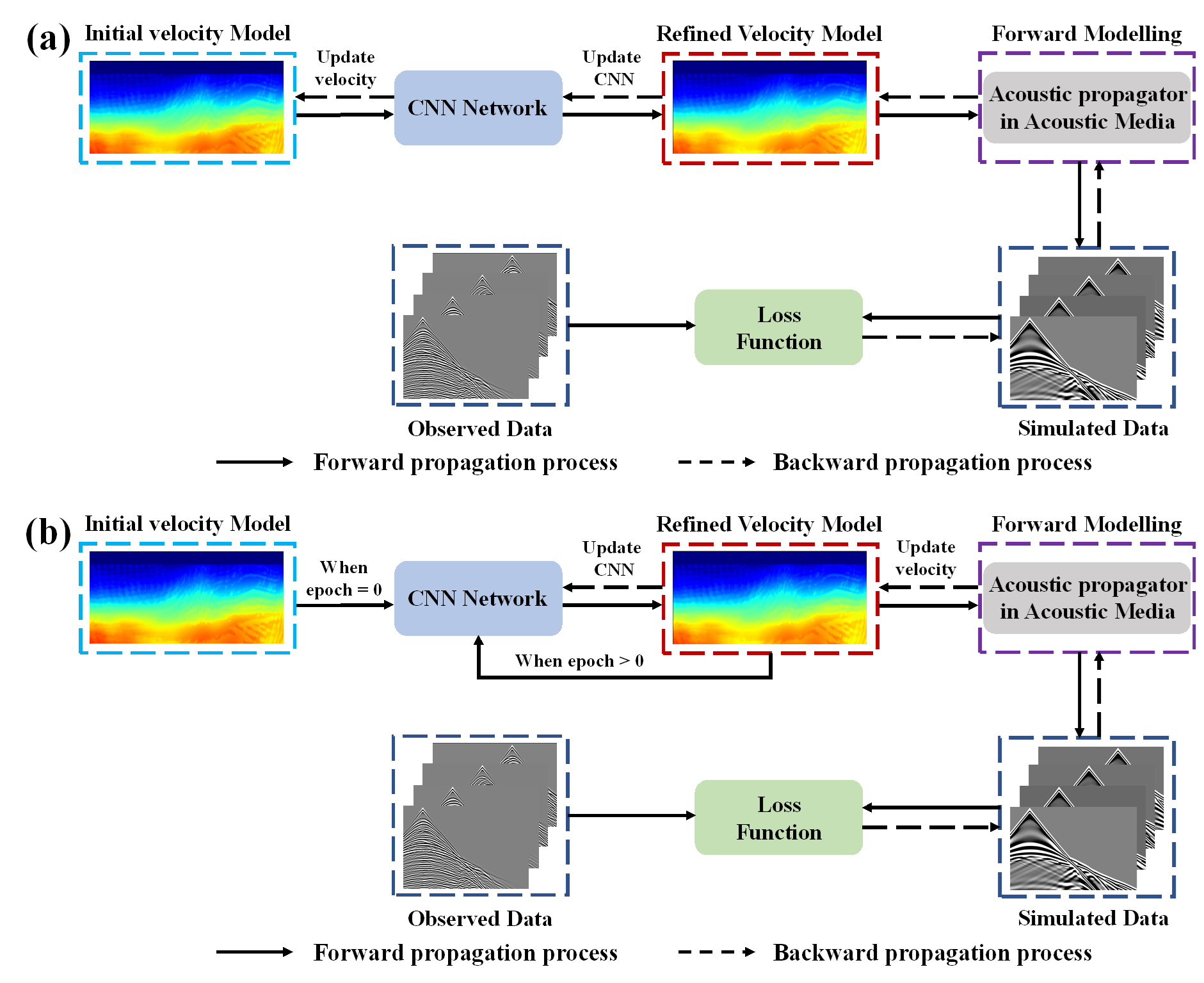}
\caption{The frameworks of the proposed VRE-FWI: The backpropagation used to update the velocity (a) passes through the CNN, and (b) does not pass through the CNN.}
\label{fig1}
\end{figure*}

We propose two different implementation frameworks for VRE-FWI, as shown in Fig. \ref{fig1}. In Framework 1 (VRE-FWI-1), a CNN is initially employed to refine the velocity model, followed by forward modeling. The residual between the observed and simulated data is then used in backpropagation to update both the velocity model and the CNN, ensuring a self-supervised learning process. We employ two Adam optimizers to update the velocity model and CNN parameters. Collectively, the velocity model and CNN form a variant of FWIDIP. The key distinction is that our approach represents the velocity using both a gridded model and a CNN, both of which are updated, whereas FWIDIP relies only on a neural network for velocity representation. When the velocity model remains unchanged and only the CNN is updated, VRE-FWI essentially becomes FWIDIP. By setting the network's weights to zero, the framework reverts to conventional FWI.

In Framework 2 (VRE-FWI-2), the velocity model is updated without passing through the CNN. The primary distinction from traditional FWI lies in an additional step where the velocity model from the previous iteration is input into the CNN for refinement. Similar to Framework 1, Framework 2 also adopts a self-supervised deep learning strategy, wherein the loss between the observed and simulated data is utilized for backpropagation to simultaneously update both the velocity model and the CNN parameters. Since the CNN in both frameworks acts as an extension of the grid-based velocity model, we refer to this approach as VRE-FWI.

\subsection{The Architecture of the CNN}
\begin{figure*}
\centering
\includegraphics[width=1\textwidth]{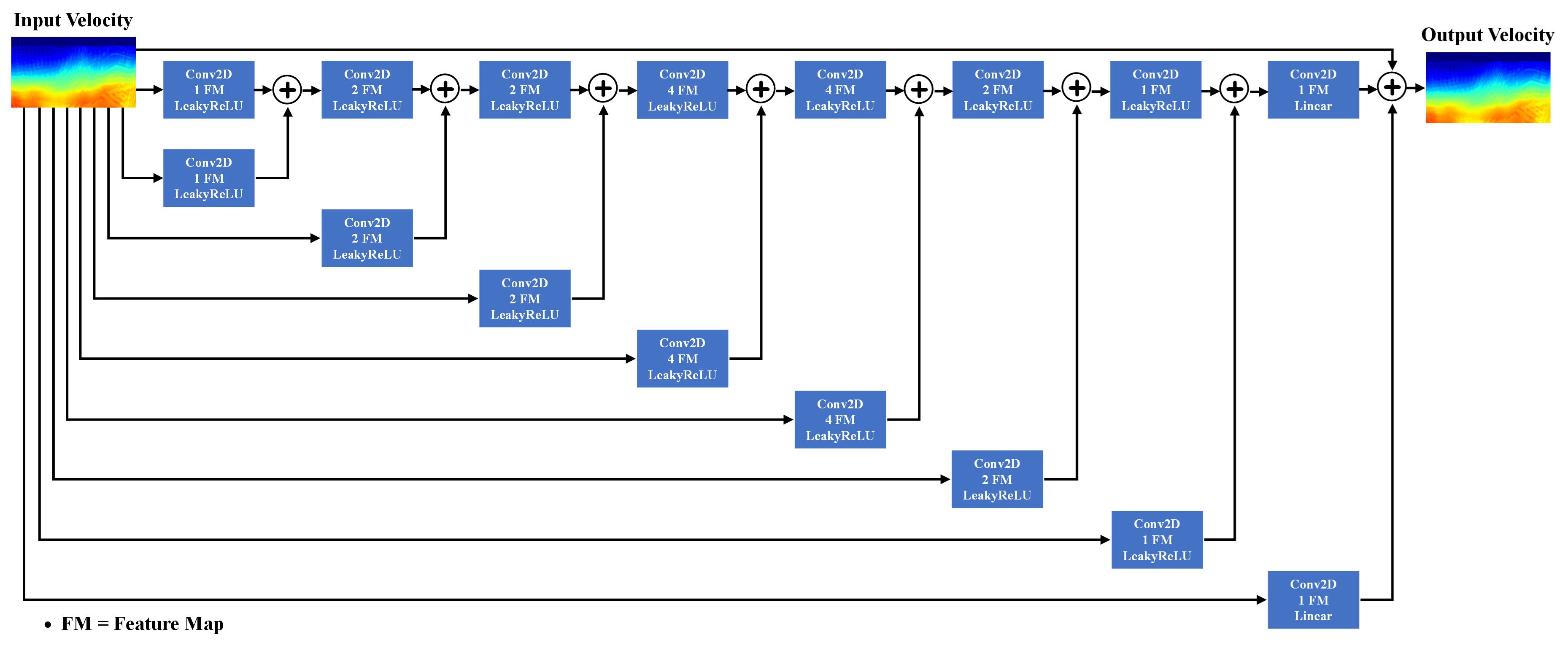}
\caption{The proposed CNN architecture for VRE-FWI.}
\label{fig2}
\end{figure*}

Fig. \ref{fig2} shows the used CNN architecture, initially introduced by \cite{saad2024siamesefwi}. The CNN has eight convolutional layers, with the number of feature maps being 1, 2, 2, 4, 4, 2, 1, and 1 from shallow to deep layers. A 3×3 convolutional kernel is used in each layer to ensure that the output size remains consistent with the input. The final layer employs a linear activation function to produce the refined velocity model, while the remaining layers utilize LeakyReLU activation to introduce non-linearity and enable the network to capture complex feature representations. To enhance the network's ability to learn complex wavefields, each layer includes a residual block that directly links the CNN's input to the output of that layer. The CNN parameters are randomly initialized and are progressively updated during the iterative FWI process. Therefore, a skip connection is implemented to directly connect the input and output of the CNN, preventing random fluctuations in the early iterations caused by random network initialization to degrade the prediction and thereby ensuring the stability of the inversion results. As the iterative process progresses, the CNN gradually learns to add features into the velocity or attenuate existing noise from it, guided by the data loss. It is worth noting that the CNN used here has been selected based on our experiments to be suitable for most model sizes and complexities. However, for larger-scale or more complex velocity models, a CNN with greater depth and width may be required.

\section{NUMERICAL EXPERIMENTS}
In this section, we present two synthetic tests that demonstrate the proposed VRE-FWI methods achieve higher inversion accuracy compared to FWI and FWIDIP at the same number of iterations. These tests also indicate that VRE-FWI incurs only a slight increase in computational cost relative to traditional FWI. Furthermore, we validate the effectiveness of VRE-FWI in enhancing inversion accuracy using a marine field dataset collected from Northwestern Australia. For the seismic wave modeling in the three tests, a free-surface boundary condition is applied at the top, while perfectly matched layer absorbing boundaries are implemented on the remaining sides to suppress boundary reflections \cite{berenger1994perfectly}.

\subsection{The SEAM model}
\begin{figure*}
\centering
\includegraphics[width=1\textwidth]{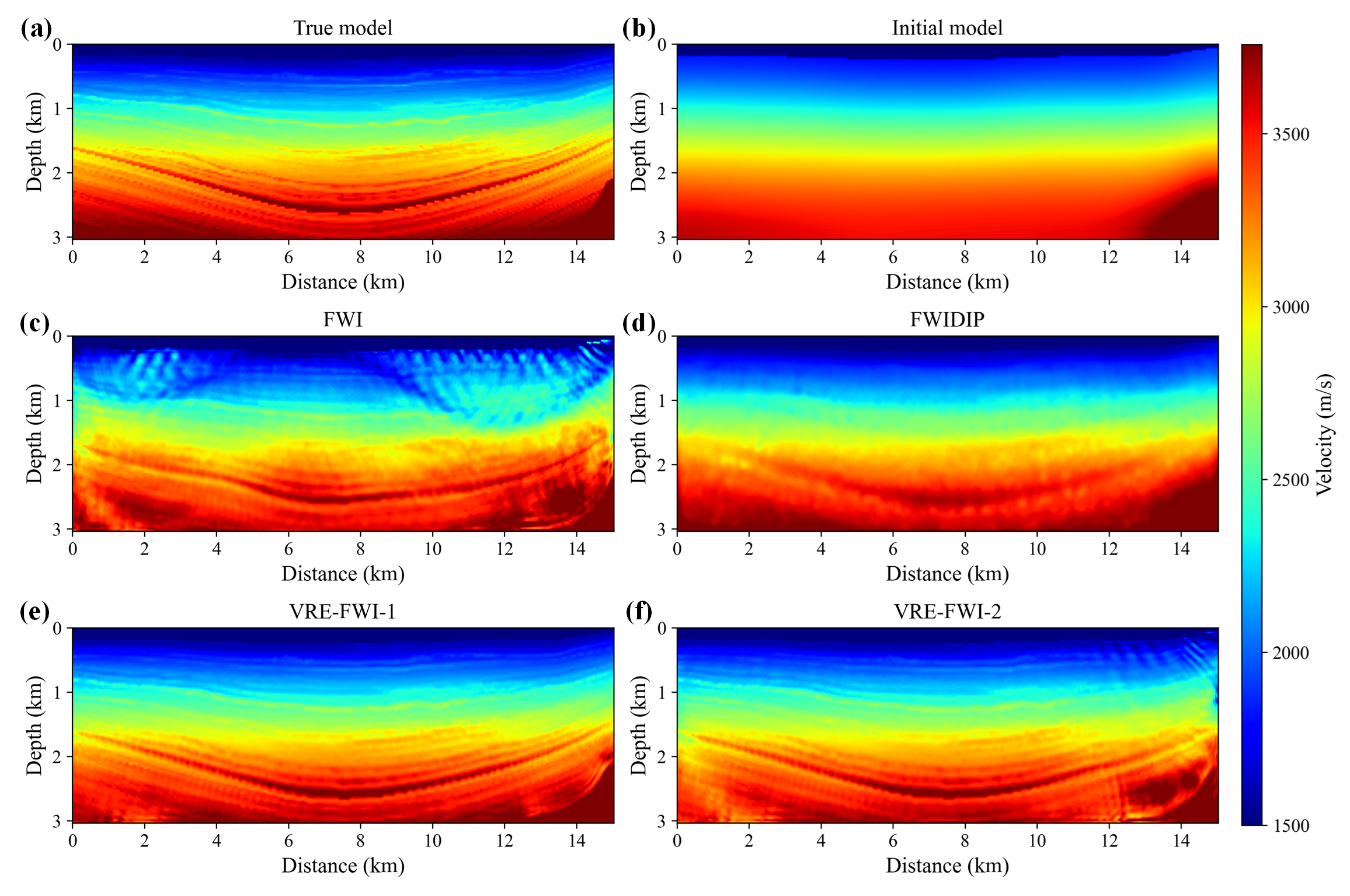}
\caption{Comparison of results for the SEAM model: (a) True SEAM model, (b) Initial velocity model, and the inverted models obtained after 100 iterations using (c) FWI, (d) FWIDIP, (e) VRE-FWI-1, and (f) VRE-FWI-2.}
\label{fig3}
\end{figure*}

Fig. \ref{fig3}(a) shows a segment of the SEAM Phase I RPSEA Model \cite{fehler2008seg}, which primarily consists of very thin stratigraphic structures. This model serves as a platform to evaluate the effectiveness of the proposed VRE-FWI in enhancing inversion accuracy for thin-layer structures compared to conventional FWI. The SEAM model used in this study has a grid size of 501×101 with a uniform grid spacing of 30 m in both the horizontal and vertical directions. For forward modeling, we employ a Ricker wavelet with a dominant frequency of 5 Hz as the source. A total of 30 shots are evenly distributed along the surface at intervals of 480 m, with the first shot positioned at the leftmost end of the model. Each shot is recorded by 501 receivers, spaced 30 m apart, and all receivers are placed at the surface. The recording duration for each shot is 6 seconds, with a time sampling interval of 2 ms. For conventional FWI, the learning rate for the velocity optimizer is set to 20. For VRE-FWI-1 and VRE-FWI-2, the learning rate for the velocity optimizer is the same as that used in FWI, while the learning rates for the CNN optimizers are set to 1e-4 and 1e-6, respectively. The initial velocity model is shown in Fig. \ref{fig3}(b), while Figs \ref{fig3}(c)–\ref{fig3}(f) show the velocity inversion results obtained using FWI, FWIDIP, VRE-FWI-1, and VRE-FWI-2, respectively, after 100 iterations. The FWI inversion result exhibits significant noise on both sides of the model, leading to lower inversion accuracy. The results from FWIDIP effectively eliminate this noise; however, due to the frequency bias of the neural network, the output appears overly smooth, making it difficult to achieve high-resolution imaging of thin layers. In contrast, the results from VRE-FWI-1 are free from noise interference while providing high-resolution velocity inversion results. VRE-FWI-2 also yields high-resolution inversion results, with only a small amount of noise remaining on the right side of the model. These findings indicate that VRE-FWI effectively reduces noise interference and achieves higher-resolution imaging compared to both FWI and FWIDIP.  

To evaluate the inversion results quantitatively, we compute the signal-to-noise ratio (SNR), the structural similarity index measure (SSIM), and the root mean square error (RMSE) between the true and inverted models obtained from different methods \cite{yang2023fwigan}, as detailed in Table \ref{tab1}. According to Table \ref{tab1}, the result obtained using VRE-FWI-1 demonstrates the highest SNR and SSIM, coupled with the lowest RMSE, indicating that VRE-FWI-1 achieves the best inversion accuracy. Furthermore, although the VRE-FWI-2 result has a lower SNR and higher RMSE compared to FWIDIP, it exhibits a higher SSIM. 
The running time analysis indicates that VRE-FWI incurs only a marginal increase in computational cost compared to traditional FWI, with the additional processing time exceeding approximately 1\%.

\begin{table}
\centering 
\begin{tabular}{@{}cccccc@{}}
\hline
Method  & SNR (dB) & SSIM   & RMSE (km/s) & FWI iterations & Running time (s) \\ \hline
FWI     & 24.61  & 0.46 & 0.05      & 100            & \textbf{237}             \\
FWIDIP  & 30.18  & 0.63 & 0.03      & 100            & 292             \\
VRE-FWI-1 & \textbf{30.85}  & \textbf{0.83} & \textbf{0.02}      & 100            & 241             \\
VRE-FWI-2 & 26.38  & 0.76 & 0.04      & 100            & 239             \\ \hline
\end{tabular}
\caption{Comparison of inversion accuracy and running time for the SEAM model. The results are computed using a single NVIDIA GPU RTX 8000, with the highest score for each metric highlighted in bold.}
\label{tab1}
\end{table}

\begin{figure*}
\centering
\includegraphics[width=0.8\textwidth]{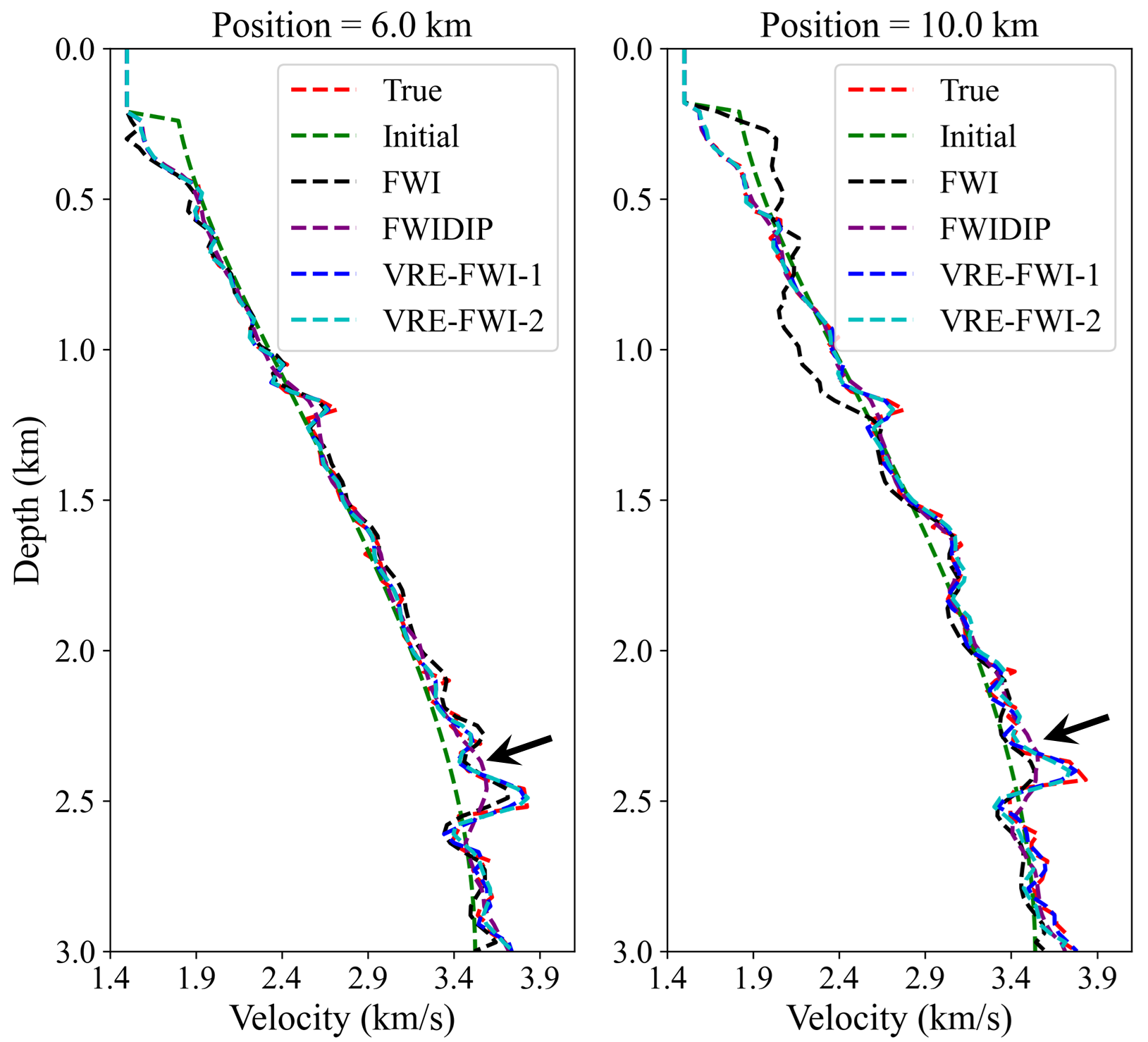}
\caption{Velocity profiles extracted from Fig. \ref{fig3} at horizontal distances of (a) 6 km and (b) 10 km, respectively.}
\label{fig4}
\end{figure*}

Fig. \ref{fig4} shows vertical velocity profiles extracted from Fig. \ref{fig3} at horizontal locations of 6 km and 10 km. As indicated by the black arrows in the Fig. \ref{fig4}, the inversion results from FWI and FWIDIP differ significantly from the true results, whereas the inverted model obtained by VRE-FWI-1 and VRE-FWI-2 closely align with the true result. This demonstrates that the VRE-FWI methods achieve higher inversion accuracy compared to FWI and FWIDIP.

\begin{figure*}
\centering
\includegraphics[width=0.6\textwidth]{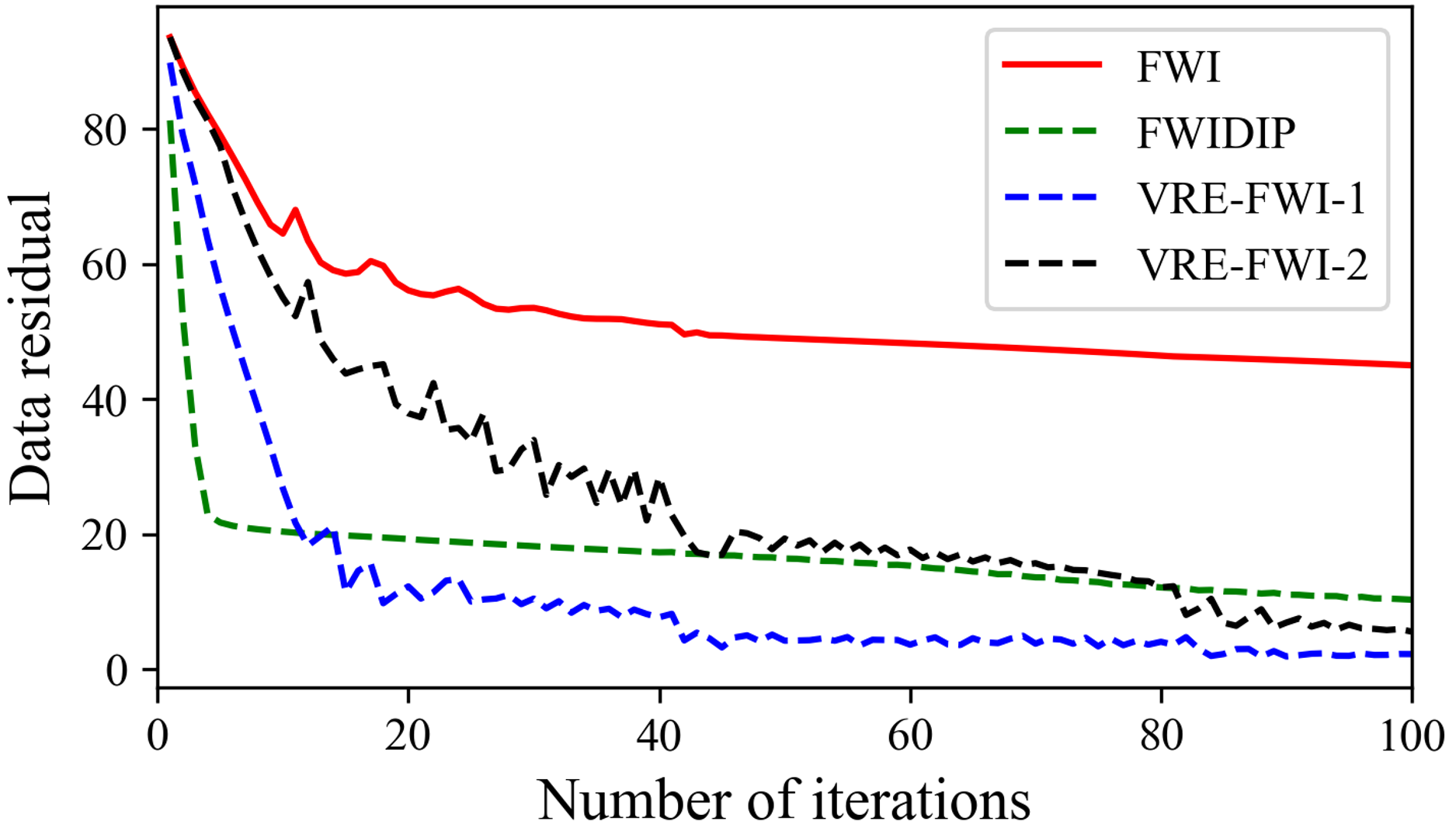}
\caption{Convergence curves of different FWI methods for the SEAM model.}
\label{fig5}
\end{figure*}

The residuals as a function of iteration number for different FWI methods are shown in Fig. \ref{fig5}. Initially, FWIDIP exhibits the fastest convergence; however, its convergence rate decreased after approximately five iterations. This is due to the low-frequency bias of the neural network; while FWIDIP effectively inverts low-wavenumber features, it struggles with the inversion of thin layers that exhibit high wavenumbers. Around the fifteenth iteration, the residual of VRE-FWI-1 begins to decrease below that of FWIDIP. Although VRE-FWI-2 shows a slower convergence rate compared to VRE-FWI-1, its final data residual is only slightly larger than that of VRE-FWI-1. Both VRE-FWI-1 and VRE-FWI-2 have data residuals that are significantly smaller than those of FWI.

\subsection{Marmousi model}
\begin{figure*}
\centering
\includegraphics[width=\textwidth]{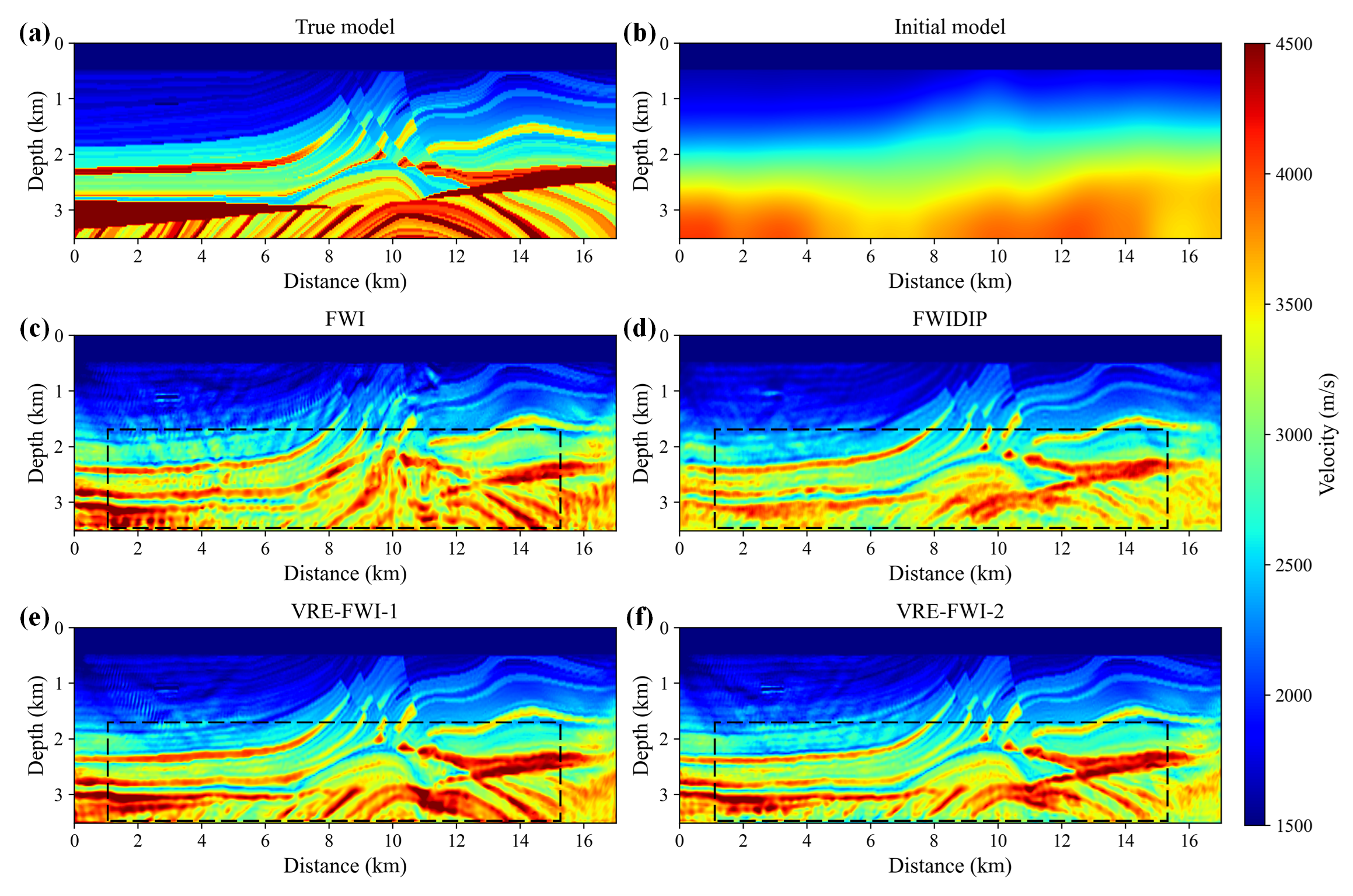}
\caption{Results comparison for the Marmousi2 model: (a) True Marmousi2 model, (b) Initial velocity model, and the inverted models from (c) FWI, (d) FWIDIP, (e) VRE-FWI-1, and (f) VRE-FWI-2 after 300 iterations.}
\label{fig6}
\end{figure*}

As shown in Fig. \ref{fig6}(a), the Marmousi2 model \cite{martin2006marmousi2} is used to evaluate the effectiveness of the proposed VRE-FWI. The model has a grid size of 567×117, with a spatial grid spacing of 30 m. 30 sources are uniformly positioned on the surface, each given by a Ricker wavelet with a dominant frequency of 5 Hz. Additionally, 567 receivers are evenly distributed along the surface. Seismic data is recorded over a duration of 6 seconds, with a time sampling interval of 3 ms. Similar to the SEAM model test, the learning rate for the velocity optimizer in FWI is set to 20. For VRE-FWI-1 and VRE-FWI-2, the velocity optimizer's learning rate is equal to that of FWI, while the learning rates for the CNN optimizers are configured to 1e-4 and 1e-6, respectively. Fig. \ref{fig6}(b) shows the initial velocity model. The inversion result obtained using conventional FWI is shown in Fig. \ref{fig6}(c). As highlighted by the black rectangular box, the resulting velocity model includes a lot of artifacts related to numerical noise and inadequate coverage. Fig. \ref{fig6}(d) presents the FWIDIP inversion result, demonstrating that FWIDIP produces a cleaner velocity model due to the neural network representation. In addition, Figs \ref{fig6}(e) and \ref{fig6}(f) show that both proposed VRE-FWI frameworks produce significantly more accurate inversion results for the complex structures in the deeper parts of the model.

Table \ref{tab2} shows the SNR, SSIM, and RMSE values comparing the velocity models from different inversion methods with the true model. Notably, VRE-FWI-1 surpasses both FWI and FWIDIP in SNR and SSIM while achieving a lower RMSE. Moreover, although VRE-FWI-2 has a slightly lower SNR and higher RMSE than FWIDIP, it attains a superior SSIM. It is also important to highlight that both VRE-FWI frameworks introduce only a minimal increase in computational cost compared to conventional FWI, with an additional cost of slightly more than 1\%.

\begin{table}
\centering 
\begin{tabular}{@{}cccccc@{}}
\hline
Method  & SNR (dB) & SSIM   & RMSE (km/s) & FWI iterations & Running time (s) \\ \hline
FWI     & 14.9305  & 0.3454 & 0.1793      & 300            & \textbf{1081}             \\
FWIDIP  & 16.1224  & 0.4875 & 0.1563      & 300            & 1122             \\
VRE-FWI-1 & \textbf{17.1628}  & \textbf{0.5771} & \textbf{0.1386}      & 300            & 1100             \\
VRE-FWI-2 & 16.0096  & 0.5068 & 0.1583      & 300            & 1092             \\ \hline
\end{tabular}
\caption{Comparison of inversion accuracy and elapsed time for the Marmousi2 model. A single NVIDIA GPU A100 is used to compute the results. We highlight in bold the best score for each measure.}
\label{tab2}
\end{table}

\begin{figure*}
\centering
\includegraphics[width=0.8\textwidth]{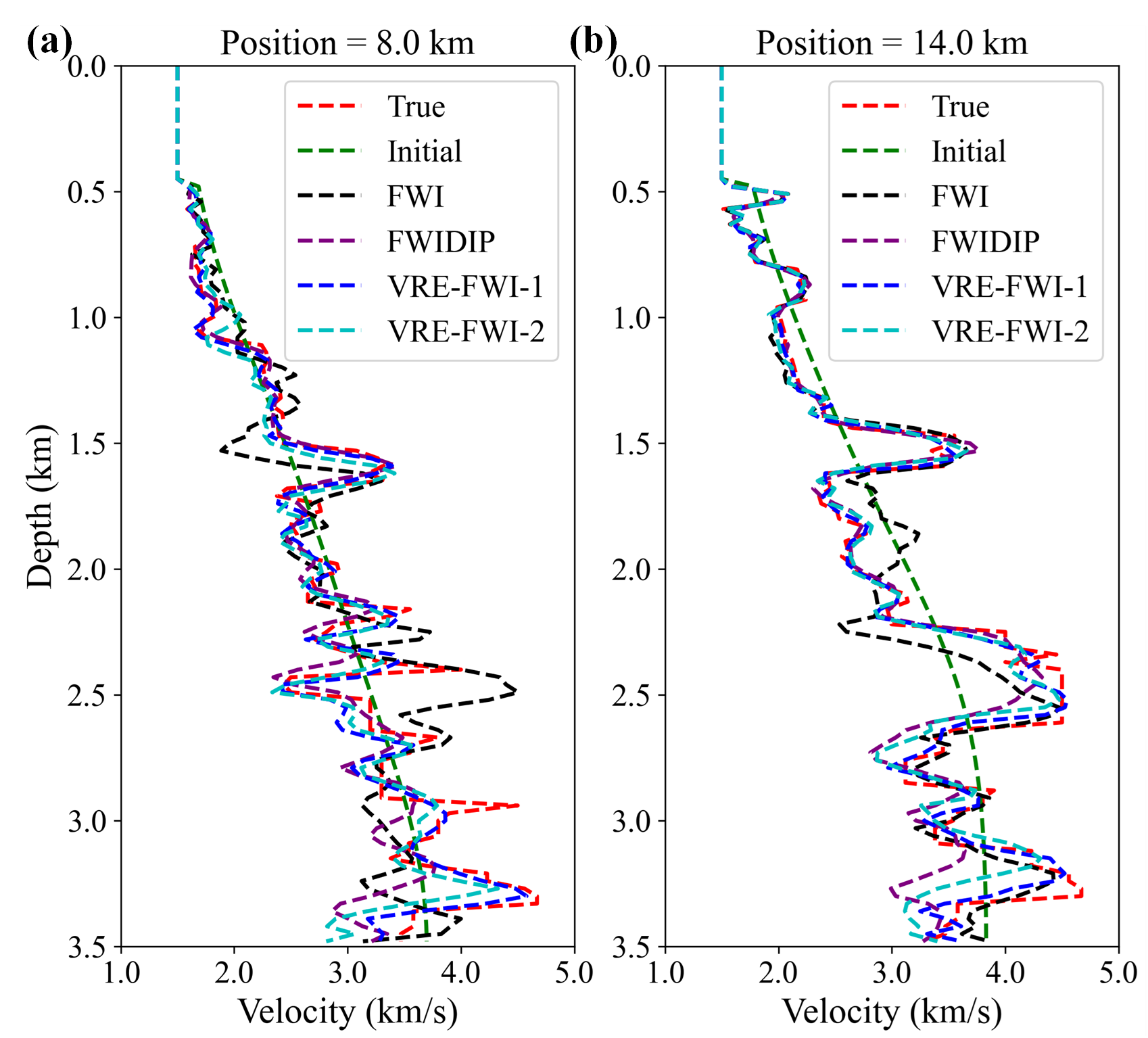}
\caption{Single trace comparison extracted from Fig. \ref{fig6} at horizontal distances of (a) 8 km and (b) 14 km.}
\label{fig7}
\end{figure*}

Fig. \ref{fig7} shows vertical velocity profile comparisons extracted from Fig. \ref{fig6} at horizontal locations 8 km and 14 km. The inversion results from FWI and FWIDIP exhibit significant discrepancies compared to the true model, particularly at depths below 2.5 km. In contrast, the results obtained using VRE-FWI-1 and VRE-FWI-2 align more closely with the true model. This demonstrates that the proposed VRE-FWI framework effectively enhances inversion accuracy compared to conventional FWI.

\begin{figure*}
\centering
\includegraphics[width=0.6\textwidth]{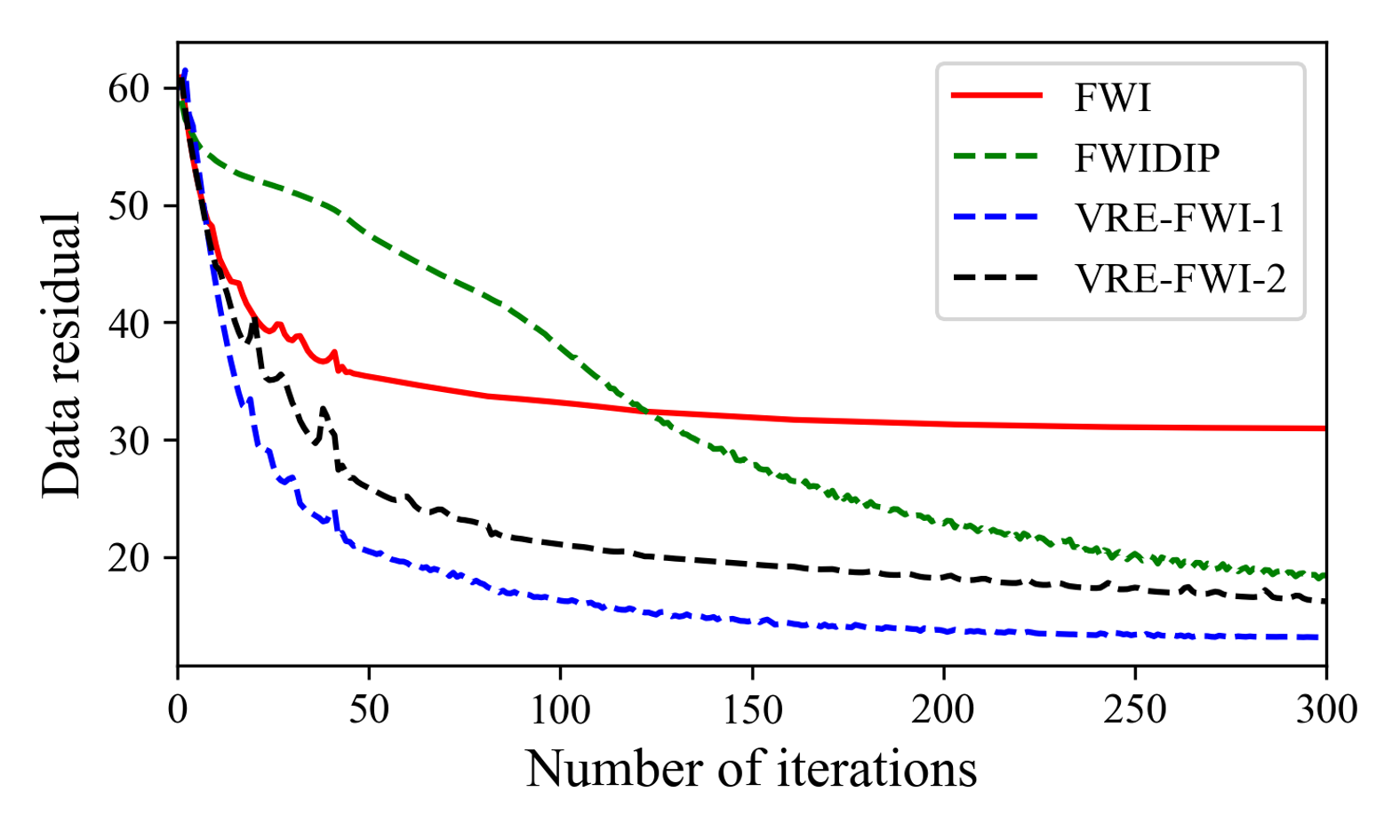}
\caption{The data residual convergence curves as a function of the number of iterations for different FWI methods applied to the Marmousi model.}
\label{fig8}
\end{figure*}

We also provide the data residual convergence curves for different methods in Fig. \ref{fig8}. From Fig. \ref{fig8}, we can observe that VRE-FWI-1 and VRE-FWI-2 achieve faster convergence rates and smaller final data residual.

\subsection{Field data}
We validate the effectiveness of the proposed VRE-FWI-1 and VRE-FWI-2 using a marine dataset acquired from North Western Australia. The dataset includes 116 shot records generated by airgun sources. It spans approximately 20 km laterally, with a shot interval of about 90 meters. For each shot, data were recorded using a towed streamer equipped with 648 hydrophones, spaced 12.5 meters apart. Each shot was recorded for a duration of 7 seconds with a time sampling interval of 1 ms. 

\begin{figure*}
\centering
\includegraphics[width=1\textwidth]{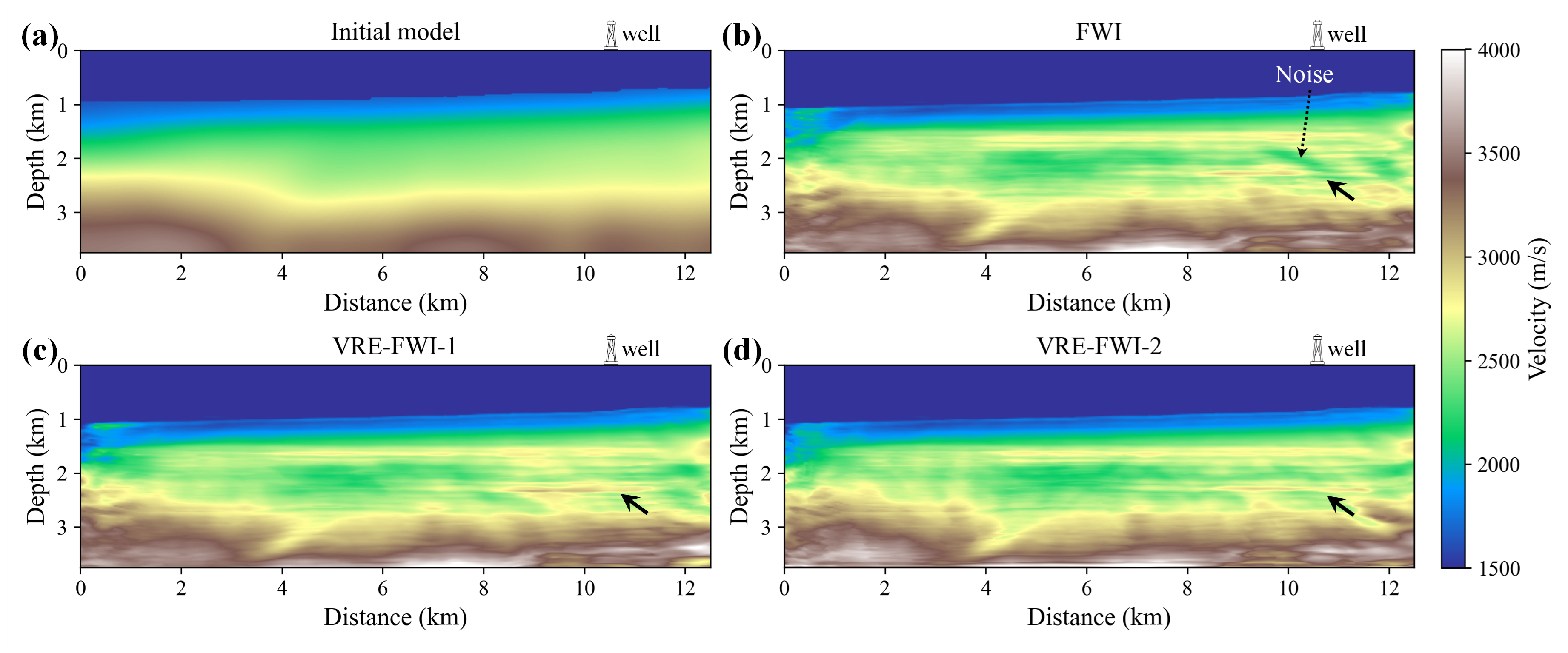}
\caption{Inversion results obtained using different FWI methods on the data collected from Western Australia: (a) initial velocity model; (b), (c), and (d) represent the results obtained using FWI, VRE-FWI-1, and VRE-FWI-2, respectively.}
\label{fig9}
\end{figure*}

Before performing FWI, we first mute samples prior to the first arrivals. To improve computational efficiency without introducing numerical dispersion, we resample the shot record time interval to 2 ms. In addition, the grid spacing of the model in the numerical simulation is set to 25 m. To further enhance inversion accuracy, we estimate the source wavelet for each shot using the known seawater velocity of 1500 m/s and the direct wave signals from near-offset traces. The initial velocity model used for FWI, shown in Fig. \ref{fig9}(a), is obtained from migration velocity analysis \cite{kalita2017efficient}. For this test, a cross-correlation loss function is employed to mitigate the impact of waveform amplitude on inversion accuracy and improve the robustness of the inversion \cite{choi2012application}.

\begin{figure*}
\centering
\includegraphics[width=0.36\textwidth]{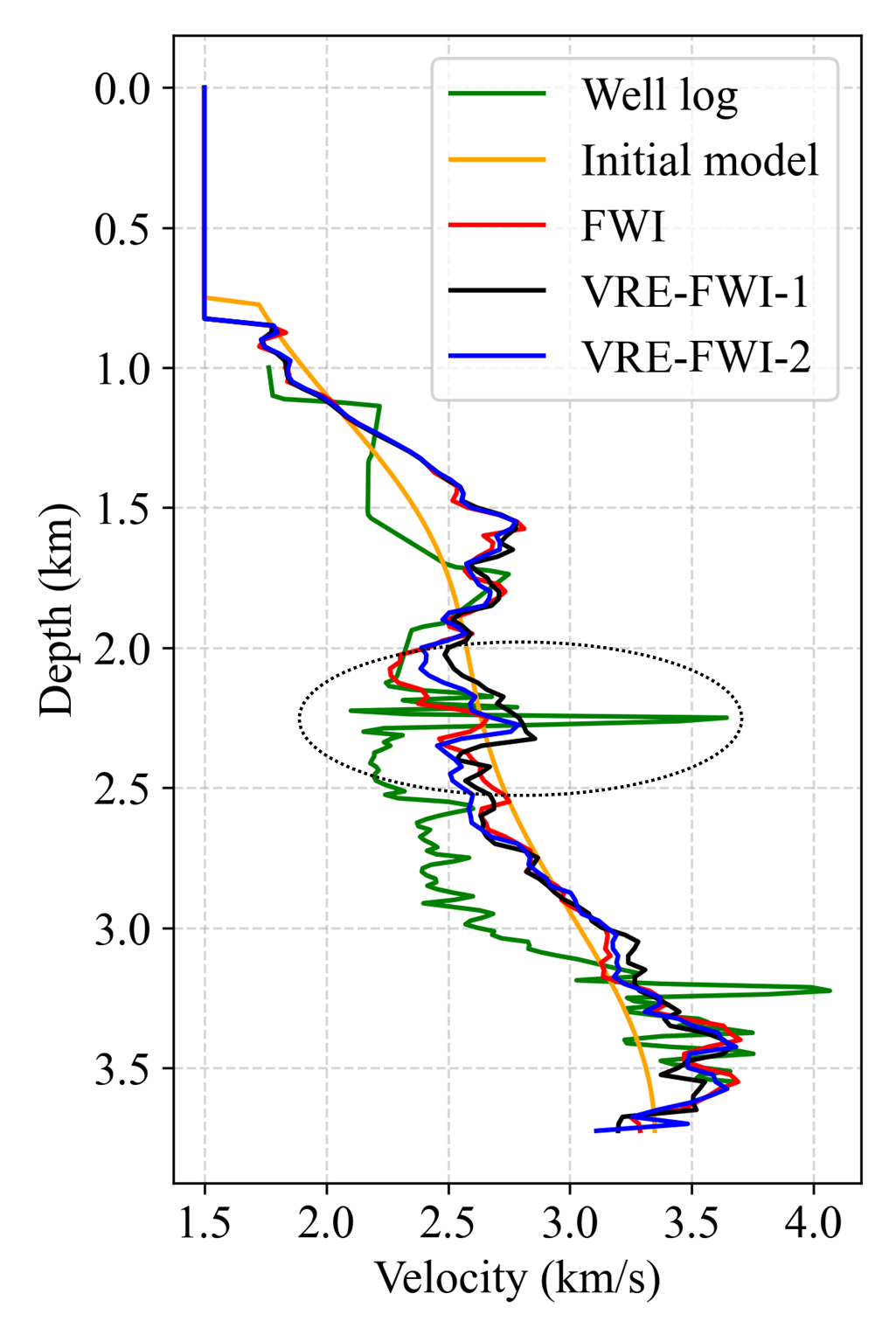}
\caption{Comparison between the well log and velocity inversion results, with the well located at 10.4 km.}
\label{fig10}
\end{figure*}

We implement the FWI process using a multi-scale inversion strategy, starting from 5 Hz, progressing through 10 Hz, 13 Hz, and up to 16 Hz. Each frequency-scale FWI stage consists of 100 iterations. The step lengths of the velocity optimizer for all three FWI methods (FWI, VRE-FWI-1, and VRE-FWI-2) are set to 10, 5, 1, and 1 for the four frequency-scale stages, respectively. For VRE-FWI-1 and VRE-FWI-2, the learning rates of the CNN parameter optimizer are set to (7e-5, 5e-6, 1e-8, 1e-9) and (5e-6, 1e-7, 3e-8, 1e-10), respectively, across the four frequency-scale stages. Figs \ref{fig9}(b)-\ref{fig9}(d) show the velocity inversion results obtained using FWI, VRE-FWI-1, and VRE-FWI-2, respectively. As indicated by the black dashed arrows in the FWI result, a low-velocity anomaly is evident, which may have been caused by noise in the gradient. However, the velocity models obtained using VRE-FWI-1 and VRE-FWI-2 effectively remove this low-velocity anomaly, resulting in a more consistent velocity structure in the horizontal direction. Additionally, as indicated by the black solid arrows, the velocity models obtained from VRE-FWI-1 and VRE-FWI-2 reveal a high-velocity structure at a depth of approximately 2.25 km, aligning well with the well log data shown in Fig. \ref{fig10}. In contrast, this high-velocity structure is disrupted by noise in the FWI results. As shown in Fig. \ref{fig10}, the inversion results of VRE-FWI-1 and VRE-FWI-2 are similar to those of FWI, except at a depth of 2.25 km, where the dashed elliptical box indicates that the VRE-FWI inversion results exhibit higher velocity values compared to the FWI results, and these values align more closely with the well log data. For the 5 Hz inversion stage, FWI, VRE-FWI-1, and VRE-FWI-2 run for 1322 s, 1348 s, and 1344 s, respectively, on an NVIDIA A100 GPU (80 GB), demonstrating that VRE-FWI enhances inversion accuracy with minimal computational overhead.

\begin{figure*}
\centering
\includegraphics[width=1\textwidth]{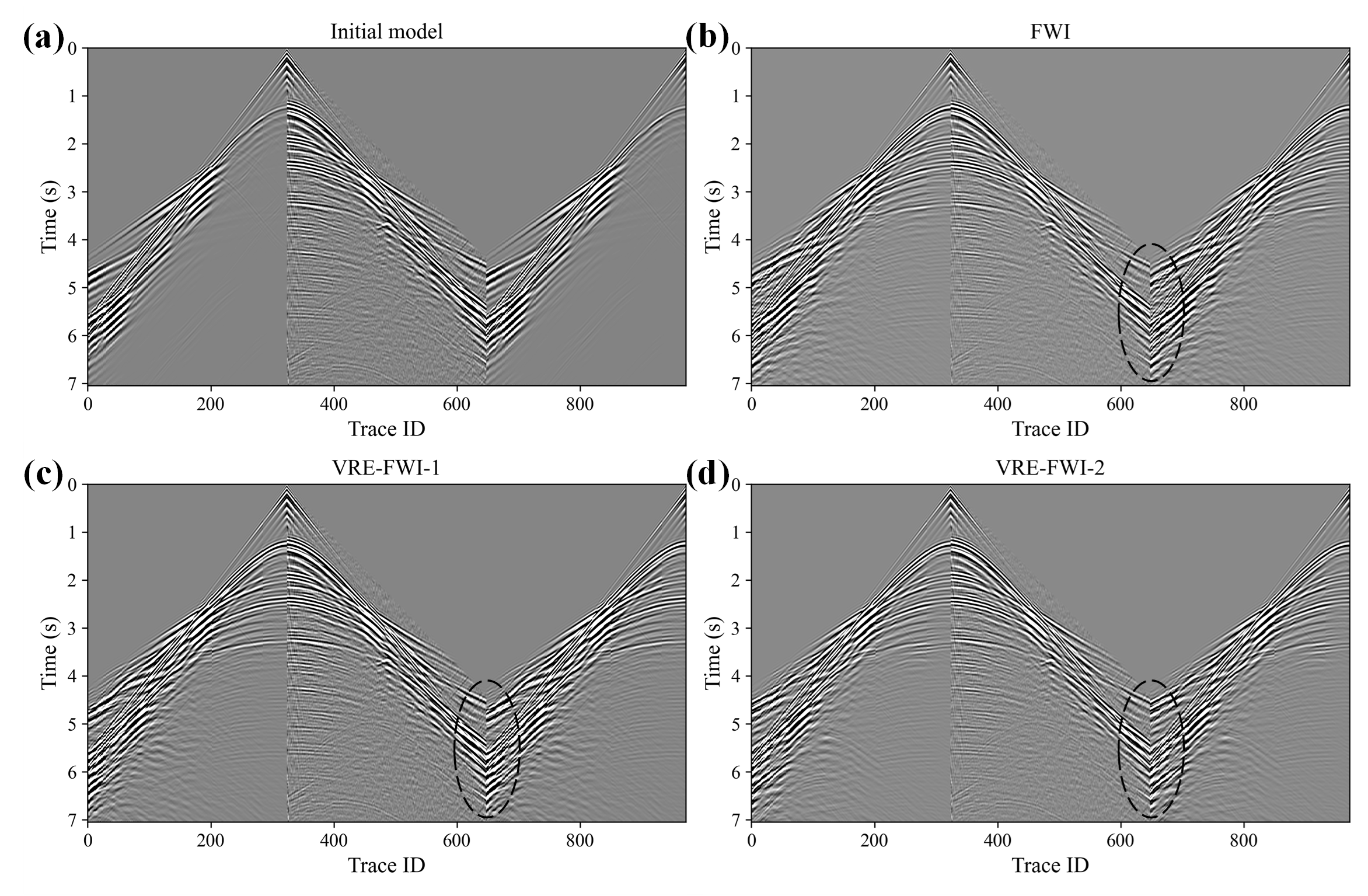}
\caption{Comparison of the observed with the simulated data produced by (a) the initial model and the inversion results from (b) FWI, (c) VRE-FWI-1, and (d) VRE-FWI-2.}
\label{fig11}
\end{figure*}

We generate shot gathers by performing forward modeling with the velocity models obtained from different inversion methods and compare them with the observed record, as shown in Fig. \ref{fig11}. By comparison, we find that the data simulated using the VRE-FWI inversion result aligns more closely with the observed data than the data simulated by the initial model and the FWI inversion result, particularly for large offsets, as indicated by the black dashed box in Fig. \ref{fig11}.

\begin{figure*}
\centering
\includegraphics[width=1\textwidth]{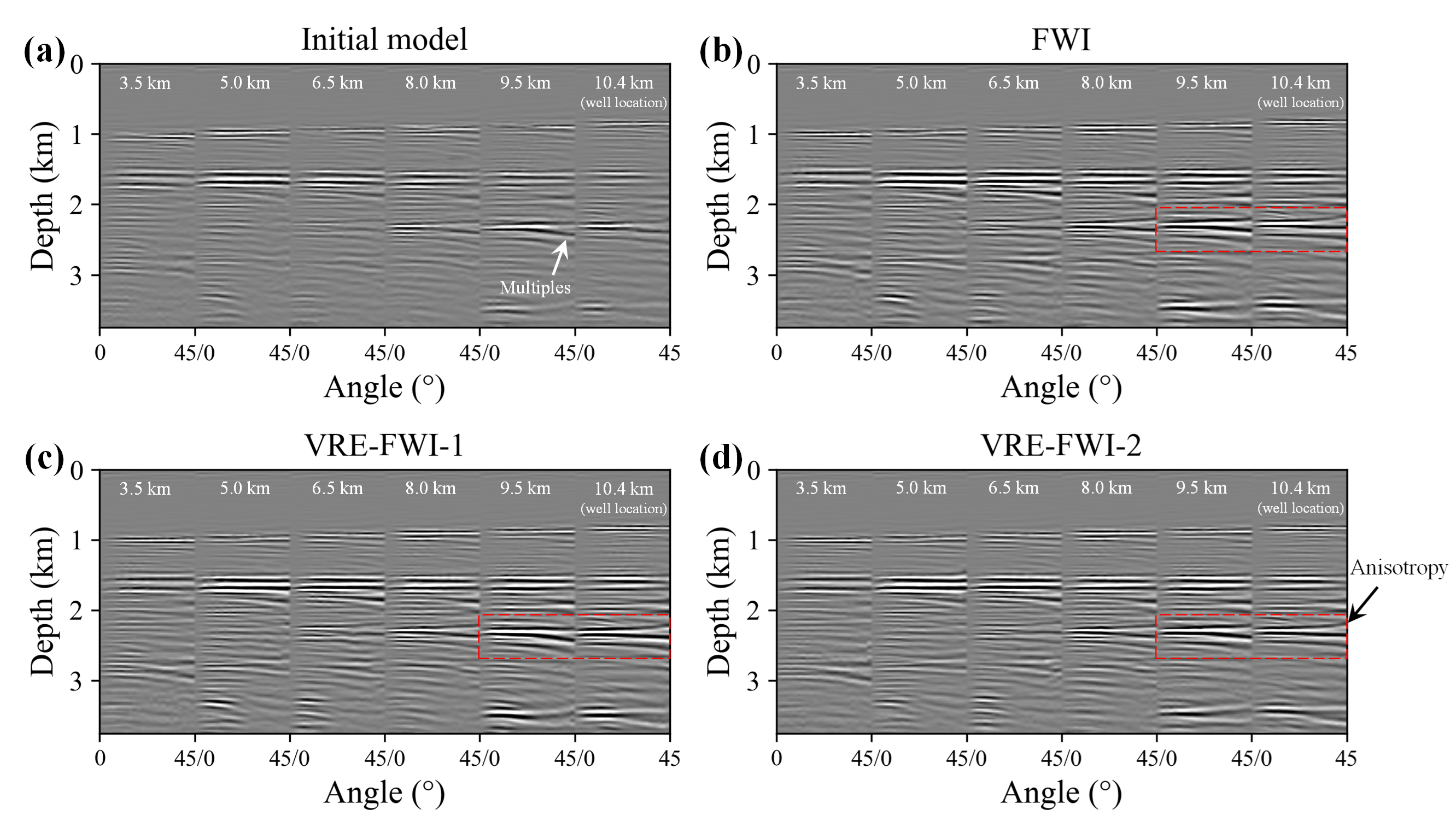}
\caption{The angle-domain common-image gathers at various positions, using (a) the initial model, (b) FWI, (c) VRE-FWI-1, and (d) VRE-FWI-2 inverted models, with each angle gather covering up to 45°.}
\label{fig12}
\end{figure*}

Fig. \ref{fig12} shows the angle-domain common-image gathers (ADCIGs) obtained using the velocity model presented in Fig. \ref{fig9}. In Fig. \ref{fig12}a, the energy of the ADCIGs obtained using the initial velocity model is relatively weak, which may be due to the inaccurate velocity model, causing the energy at the image points to be poorly focused. The downward-curved angle-domain image gather, marked by the white arrow, corresponds to multiples. The ADCIGs computed using the FWI velocity are mostly flat. However, upward-curved gathers appear at 9.5 km and 10.4 km (i.e., the well location), as highlighted by the red rectangles. In the ADCIGs computed with the VRE-FWI-1 velocity model, the gather at 9.5 km is flat, indicating that VRE-FWI-1 provides a more accurate velocity model compared to FWI. Even at 10.4 km, the gathers obtained using both the VRE-FWI-1 and VRE-FWI-2 velocity models remain upward-curved, yet they demonstrate better energy focusing compared to those from FWI. By examining the inverted velocity model shown in Fig. \ref{fig9}, we observe that the strata in this region are predominantly horizontal, suggesting that wave propagation may be influenced by vertical transverse isotropy (VTI), which is common in layered sedimentary formations. Therefore, we attribute the upward curvature of the gathers at 10.4 km to anisotropic effects. In the future, we aim to develop a VTI FWI algorithm to further enhance inversion accuracy.

\section{DISCUSSION}

\subsection{Interpretation of the function of the CNN}
\begin{figure*}
\centering
\includegraphics[width=1\textwidth]{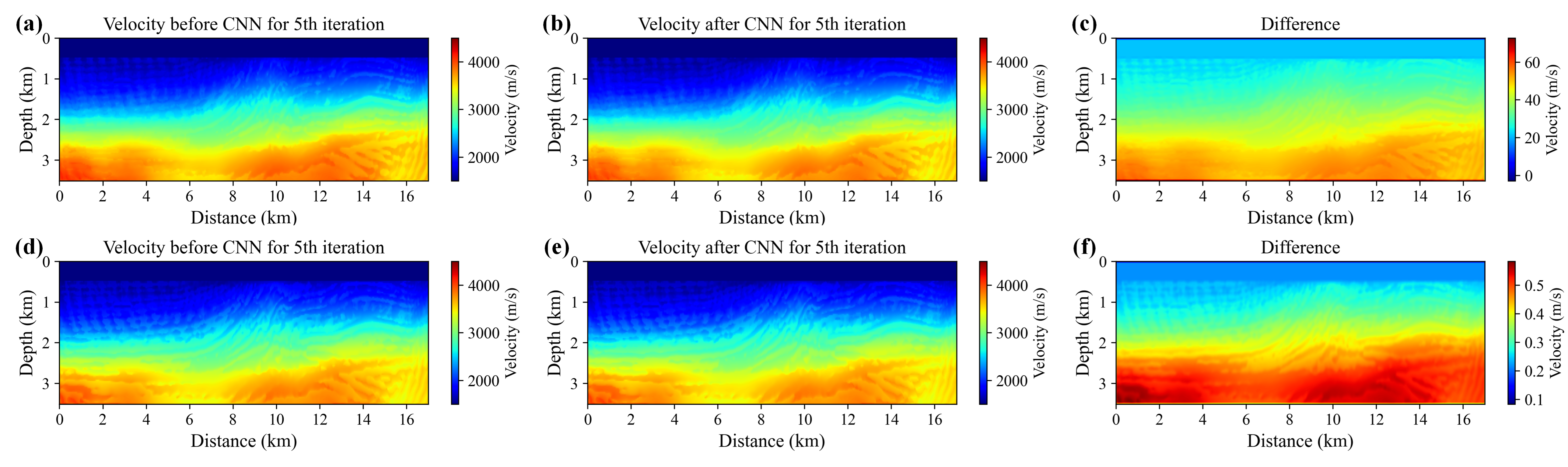}
\caption{Velocity models (a and d) before and (b and e) after CNN optimization at the 5th iteration, along with their (c and f) differences. The first and second rows show the results for VRE-FWI-1 and VRE-FWI-2, respectively. The difference is obtained by subtracting the velocity after CNN optimization from the velocity before CNN optimization.}
\label{fig13}
\end{figure*}

To illustrate the role of the CNN in VRE-FWI, we use the Marmousi model as an example and plot the velocity models before and after CNN optimization at the fifth iteration, along with their differences, as shown in Fig. \ref{fig13}. The difference is obtained by subtracting the velocity after the CNN optimization from the velocity before the optimization. As shown in Fig. \ref{fig13}, this difference yields positive values, indicating that the CNN primarily removes certain noise components, thereby optimizing the direction of the velocity updates. This is also consistent with our VRE-FWI framework and CNN architecture. Since our CNN architecture includes a skip connection that directly links the input and output, the residual before and after the CNN is essentially noise. Moreover, the difference value in VRE-FWI-1 is greater than that in VRE-FWI-2, indicating that the CNN in VRE-FWI-1 has a more pronounced effect on velocity changes. This also explains why VRE-FWI-1 converges faster than VRE-FWI-2. 

\subsection{The impact of increasing the number of epochs}
\begin{figure*}
\centering
\includegraphics[width=1\textwidth]{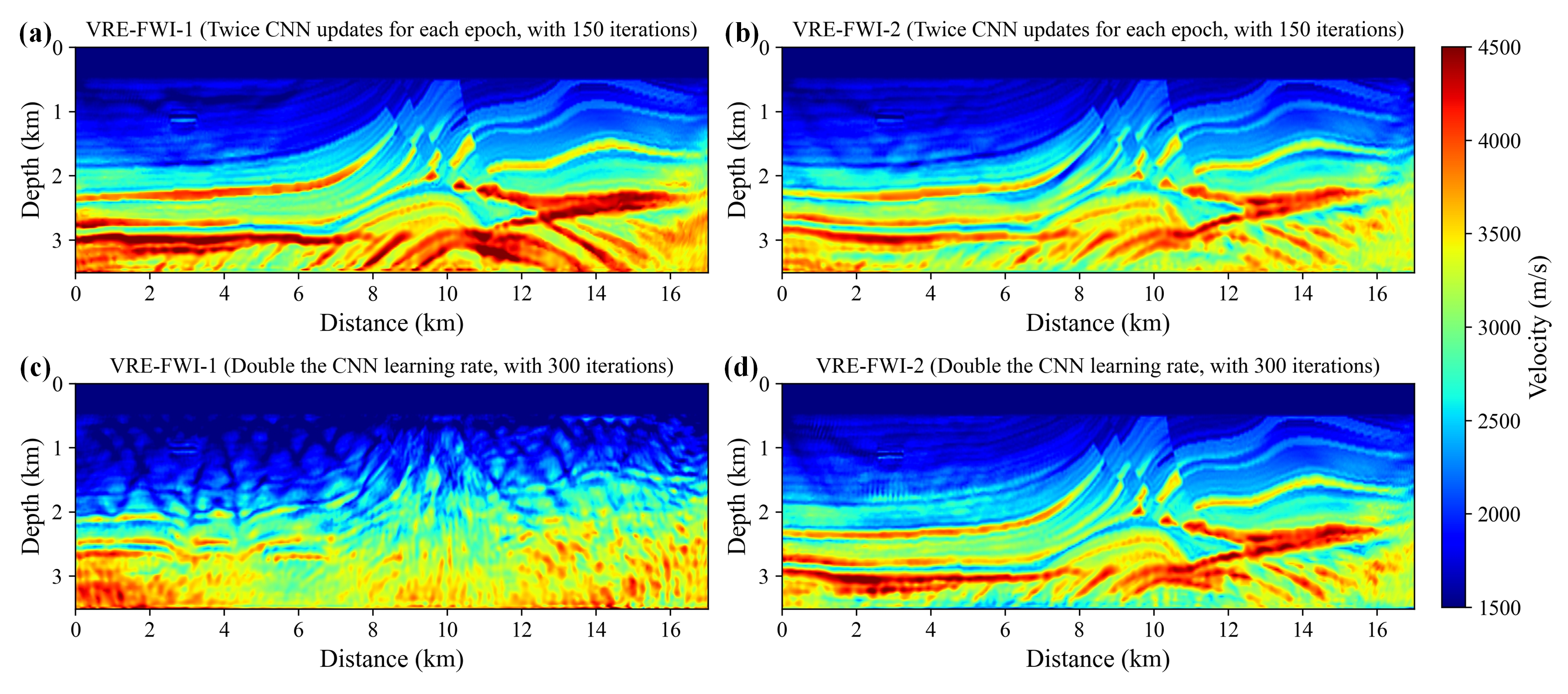}
\caption{The first row shows the inversion results of (a) VRE-FWI-1 and (b) VRE-FWI-2 after 150 iterations, where the CNN is updated twice per epoch. The second row shows the inversion results of (c) VRE-FWI-1 and (d) VRE-FWI-2 after 300 iterations, with the learning rate set to twice that used in Fig. \ref{fig6}.}
\label{fig14}
\end{figure*}

In the VRE-FWI implementation framework shown in Fig. \ref{fig1}, we update both the velocity model and the CNN once per epoch. Since the CNN is updated progressively, its ability to refine the velocity model improves with an increasing number of iterations. Therefore, we consider updating the CNN twice per epoch while still updating the velocity only once. Specifically, in each epoch, we perform the forward and backward propagation processes twice to update the CNN, while the velocity model is updated only during the second backward propagation. This leads to a doubling of the computational cost for each epoch compared to the original implementation. For testing, we use the Marmousi model, and to ensure a fair comparison of running time with Fig. \ref{fig6}, we limit the number of iterations to 150. The obtained inversion results are shown in Figs \ref{fig14}(a) and \ref{fig14}(b). Compared to the results shown in Figs \ref{fig6}(e) and \ref{fig6}(f), the results presented in Figs \ref{fig14}(a) and \ref{fig14}(b) demonstrate higher accuracy. The SNR, SSIM, and RMSE for Fig. \ref{fig14}(a) are 19.46, 0.66, and 0.10, respectively. The result in Fig. \ref{fig14}(a) exhibits higher SNR and SSIM values than those in Fig. \ref{fig6}(e), and showing a smaller RMSE. The SNR, SSIM, and RMSE values for Fig. \ref{fig14}(b) are 16.28, 0.55, and 0.15, respectively. Similarly, the result in Fig. \ref{fig14}(b) shows improved SNR and SSIM compared to those in Fig. \ref{fig6}(f), along with a smaller RMSE. Additionally, we attempt to update the CNN three times per epoch but found no improvement in inversion accuracy. Therefore,  it is concluded that updating the CNN twice per epoch strikes the best balance between inversion accuracy and computational efficiency.

We also double the learning rate of the CNN optimizer, and the inversion results for VRE-FWI-1 and VRE-FWI-2 after 300 iterations are shown in Figs \ref{fig14}(c) and \ref{fig14}(d). The inversion result of VRE-FWI-1 shows a noticeable decrease in accuracy compared to Fig. \ref{fig6}(e). The result shown in Fig. \ref{fig14}(d) has SNR, SSIM, and RMSE values of 15.69, 0.52, and 0.16, respectively. Compared to Fig. \ref{fig6}(f), both the SNR and SSIM are lower, and the RMSE is higher. This finding indicates that increasing the learning rate of the CNN does not enhance inversion accuracy and may even reduce it. 

\subsection{The sensitivity of the VRE-FWI to the noise}
\begin{figure*}
\centering
\includegraphics[width=1\textwidth]{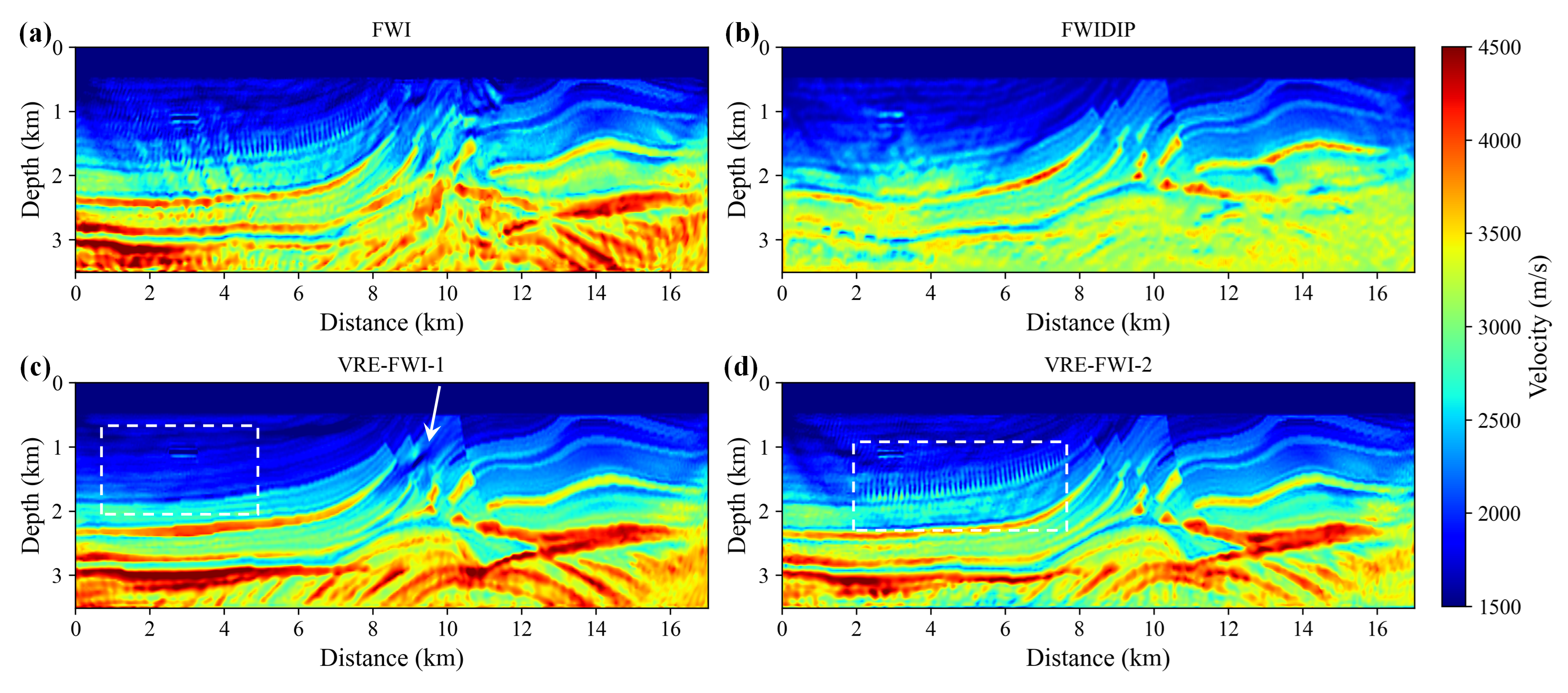}
\caption{Inversion results obtained from noisy observed records using (a) FWI, (b) FWIDIP, (c) VRE-FWI-1, and (d) VRE-FWI-2. The used observed records have a SNR of 5 Hz.}
\label{fig15}
\end{figure*}

To evaluate the robustness of VRE-FWI in handling noisy data, we use the Marmousi model as an example and add white Gaussian noise to the observed data, resulting in a SNR of 5 dB. Fig. \ref{fig15} shows the inversion results obtained using FWI, FWIDIP, VRE-FWI-1, and VRE-FWI-2. Compared to the result shown in Fig. \ref{fig6}(d), the accuracy of the FWIDIP inversion result (Fig. \ref{fig15}b) is significantly reduced. In comparison to Fig. \ref{fig6}(e), the inversion result of VRE-FWI-1 (Fig. \ref{fig15}c) shows some improvement, as indicated by the white dashed box in Fig. \ref{fig15}(c); however, there is a decrease in accuracy in the area marked by the white arrow. Compared to Fig. \ref{fig6}(f), the inversion result of VRE-FWI-2 (Fig. \ref{fig15}d) exhibits a slight reduction in accuracy, as highlighted by the white dashed box in Fig. \ref{fig15}(d). This indicates that VRE-FWI exhibits a degree of robustness against noise and remains effective even in the presence of noisy data. 

\subsection{Comparison to the FWIDIP and future work}
In this work, we introduce a novel combined representation of the velocity model that integrates a conventional gridded representation with a CNN network. This design allows for more degrees of freedom in the velocity representation, to attenuate numerical noise and refine the velocity model. Unlike FWIDIP, which relies exclusively on a neural network (updates only the neural network parameters), our method incorporates both a gridded model and a CNN for velocity representation. When the velocity model is fixed and only the CNN is adjusted, VRE-FWI effectively functions as FWIDIP. Moreover, by setting the weights of the network to zero, the framework reverts to conventional FWI. Numerical experiments demonstrate that the proposed VRE-FWI converges faster than FWIDIP due to the simultaneous update of the velocity model and CNN. Additionally, FWIDIP can invert from low to high wavenumbers, which helps alleviate local minima. In future research, we aim to leverage FWIDIP's advantage in mitigating local minima while reducing the thousands of iterations typically required. We will explore how to achieve this within the framework of VRE-FWI by fine-tuning the weights involved in the updates of both the velocity model and the CNN.

\section{Conclusion}
We proposed a velocity representation extension full waveform inversion (VRE-FWI) by using a convolutional neural network (CNN) as a supplement to the grid-based velocity model. In this approach, the CNN is employed to refine the velocity model, and both the velocity model and neural network parameters are simultaneously updated by minimizing the discrepancy between simulated and observed data, ensuring a self-supervised learning process. Due to the adaptive smoothing effect of the CNN, it effectively removes noise in the gradient caused by numerical simulation approximation and insufficient data sampling coverage, thus improving the accuracy of FWI. Tests on synthetic and real data demonstrate that, compared to conventional FWI, the proposed VRE-FWI effectively suppresses gradient noise and yields improved inversion accuracy. Compared to FWI, VRE-FWI achieves improved inversion accuracy with only a slight increase in computational cost.

\section{Acknowledgments}
The authors sincerely appreciate the support from KAUST and the DeepWave Consortium sponsors. They also thank the SWAG group for fostering a collaborative research environment. The authors gratefully acknowledge the Supercomputing Laboratory at KAUST for providing the computational resources used in this work. Special thanks are extended to Jianping Huang and Qinglin Xie from China University of Petroleum (East China) for their valuable help.

\bibliography{references.bib}
\bibliographystyle{unsrt} 






\end{document}